\documentclass[runningheads]{llncs}

\usepackage{cite}
\usepackage{graphicx}
\usepackage{amsfonts}
\usepackage{amsmath}
\usepackage{amssymb}
\usepackage{booktabs}
\usepackage{tabularx}
\usepackage{wrapfig}
\usepackage[dvipsnames]{xcolor}

\usepackage[pagebackref,breaklinks,colorlinks]{hyperref}

\usepackage[capitalize]{cleveref}
\crefname{section}{Sec.}{Secs.}
\Crefname{section}{Section}{Sections}
\Crefname{table}{Table}{Tables}
\crefname{table}{Tab.}{Tabs.}

\begin{document}

\title{NASDM: Nuclei-Aware Semantic Histopathology Image Generation Using Diffusion Models}

\titlerunning{NASDM: Nuclei-Aware Histopathology Image Generation}

\author{Aman Shrivastava,
P. Thomas Fletcher}

\authorrunning{A. Shrivastava et al.}

\institute{University of Virginia, Charlottesville, Virginia, USA \\
\email{\{as3ek,ptf8v\}@virginia.edu}}





\maketitle

\begin{abstract}
In recent years, computational pathology has seen tremendous progress driven by deep learning methods in segmentation and classification tasks aiding prognostic and diagnostic settings. Nuclei segmentation, for instance, is an important task for diagnosing different cancers. However, training deep learning models for nuclei segmentation requires large amounts of annotated data, which is expensive to collect and label. This necessitates explorations into generative modeling of histopathological images. In this work, we use recent advances in conditional diffusion modeling to formulate a first-of-its-kind nuclei-aware semantic tissue generation framework (NASDM) which can synthesize realistic tissue samples given a semantic instance mask of up to six different nuclei types, enabling pixel-perfect nuclei localization in generated samples. These synthetic images are useful in applications in pathology pedagogy, validation of models, and supplementation of existing nuclei segmentation datasets. We demonstrate that NASDM is able to synthesize high-quality histopathology images of the colon with superior quality and semantic controllability over existing generative methods.

\end{abstract}

\section{Introduction}
\label{sec:introduction}
Histopathology relies on hematoxylin and eosin (H\&E) stained biopsies for microscopic inspection to identify visual evidence of diseases. Hematoxylin has a deep blue-purple color and stains acidic structures such as DNA in cell nuclei. Eosin, alternatively, is red-pink and stains nonspecific proteins in the cytoplasm and the stromal matrix. Pathologists then examine highlighted tissue characteristics to diagnose diseases, including different cancers. A correct diagnosis, therefore, is dependent on the pathologist's training and prior exposure to a wide variety of disease subtypes~\cite{xie2020integrating}. This presents a challenge, as some disease variants are extremely rare, making visual identification difficult. In recent years, deep learning methods have aimed to alleviate this problem by designing discriminative frameworks that aid diagnosis~\cite{van2021deep, wu2022recent}. Segmentation models find applications in spatial identification of different nuclei types~\cite{graham2019hover} or directly detecting visual aberrations like breast cancer metastasis.

However, generative modeling in histopathology is relatively unexplored. Generative models can generate realistic synthetic images unconditionally or given a conditioning signal. They can be used to generate histopathology images with specific characteristics, such as visual patterns identifying rare cancer subtypes~\cite{fajardo2021oversampling}. As such, generative models can be sampled to emphasize each disease subtype equally and generate more balanced datasets, thus preventing dataset biases getting amplified by the models~\cite{hall2022systematic}. Generative models have the potential to improve the pedagogy, trustworthiness, generalization, and coverage of disease diagnosis in the field of histology by aiding both deep learning models and human pathologists. Synthetic datasets can also tackle privacy concerns surrounding medical data sharing. Additionally, conditional generation of annotated data adds even further value to the proposition as labeling medical images involves tremendous time, labor, and training costs. Recently, denoising diffusion probabilistic models (DDPMs)~\cite{ho2020denoising} have achieved tremendous success in conditional and unconditional generation of real-world images~\cite{dhariwal2021diffusion}. Further, the semantic diffusion model (SDM) demonstrated the use of DDPMs for generating images given semantic layout~\cite{wang2022semantic}. In this work, (1) we leverage recently discovered capabilities of DDPMs to design a first-of-its-kind nuclei-aware semantic diffusion model (NASDM) that can generate realistic tissue patches given a semantic mask comprising of multiple nuclei types, (2) we train our framework on the Lizard dataset~\cite{graham2021lizard} consisting of colon histology images and achieve state-of-the-art generation capabilities, and (3) we perform extensive ablative, qualitative, and quantitative analyses to establish the proficiency of our framework on this tissue generation task.

\section{Related Work}
\label{sec:background}
Deep learning based generative models for histopathology images have seen tremendous progress in recent years due to advances in digital pathology, compute power, and neural network architectures. Several GAN-based generative models have been proposed to generate histology patches~\cite{levine2020synthesis, xue2021selective, zhou2022u}. However, GANs suffer from problems of frequent mode collapse and overfitting their discriminator~\cite{xiao2021tackling}. It is also challenging to capture long-tailed distributions and synthesize rare samples from imbalanced datasets using GANs. More recently, denoising diffusion models have been shown to generate highly compelling images by incrementally adding information to noise~\cite{ho2020denoising}. Success of diffusion models in generating realistic images led to various conditional~\cite{ kawar2022denoising, saharia2022palette, saharia2022image} and unconditional~\cite{dhariwal2021diffusion, ho2022cascaded, nichol2021improved} diffusion models that generate realistic samples with high fidelity. Following this, a morphology-focused diffusion model has been presented for generating tissue patches based on genotype~\cite{moghadam2023morphology}. Semantic image synthesis is a task involving generating diverse realistic images from semantic layouts. GAN-based semantic image synthesis works~\cite{tan2021diverse, tan2021efficient, park2019semantic} generally struggled at generating high quality and enforcing semantic correspondence at the same time. To this end, a semantic diffusion model has been proposed that uses conditional denoising diffusion probabilistic model and achieves both better fidelity and diversity~\cite{wang2022semantic}. We use this progress in the field of conditional diffusion models and semantic image synthesis to formulate our NASDM framework.


\section{Method}
\label{sec:method}
In this paper, we describe our framework for generating tissue patches conditioned on semantic layouts of nuclei. Given a nuclei segmentation mask, we intend to generate realistic synthetic patches. In this section, we (1) describe our data preparation, (2) detail our stain-normalization strategy, (3) review conditional denoising diffusion probabilistic models, (4) outline the network architecture used to condition on semantic label map, and (5) highlight the classifier-free guidance mechanism that we employ at sampling time. 

\subsection{Data Processing} \label{sec:data_process}
We use the Lizard dataset~\cite{graham2021lizard} to demonstrate our framework. This dataset consists of histology image regions of colon tissue from six different data sources at $20\times$ objective magnification. The images are accompanied by full segmentation annotation for different types of nuclei, namely, epithelial cells, connective tissue cells, lymphocytes, plasma cells, neutrophils, and eosinophils. A generative model trained on this dataset can be used to effectively synthesize the colonic tumor micro-environments. The dataset contains $238$ image regions, with an average size of $1055\times934$ pixels. As there are substantial visual variations across images, we construct a representative test set by randomly sampling a 7.5\% area from each image and its corresponding mask to be held-out for testing. The test and train image regions are further divided into smaller image patches of $128\times128$ pixels at two different objective magnifications: (1) at $20\times$, the images are directly split into $128\times128$ pixels patches, whereas (2) at $10\times$, we generate $256\times256$ patches and resize them to $128\times128$ for training. To use the data exhaustively, patching is performed with a $50\%$ overlap in neighboring patches. As such, at (1) $20\times$ we extract a total of 54,735 patches for training and 4,991 patches as a held-out set, while at (2) $20\times$ magnification we generate 12,409 training patches and 655 patches are held out.

\subsection{Stain Normalization}
A common issue in deep learning with H\&E stained histopathology slides is the visual bias introduced by variations in the staining protocol and the raw materials of chemicals leading to different colors across slides prepared at different labs~\cite{bejnordi2014quantitative}. As such, several stain-normalization methods have been proposed to tackle this issue by normalizing all the tissue samples to mimic the stain distribution of a given target slide~\cite{macenko2009method, vahadane2016structure, shrivastava2021self}. In this work, we use the structure preserving color normalization scheme introduce by Vahadane et al.~\cite{vahadane2016structure} to transform all the slides to match the stain distribution of an empirically chosen slide from the training dataset.


\begin{figure}[t]
\begin{center}
\includegraphics[width=\linewidth]{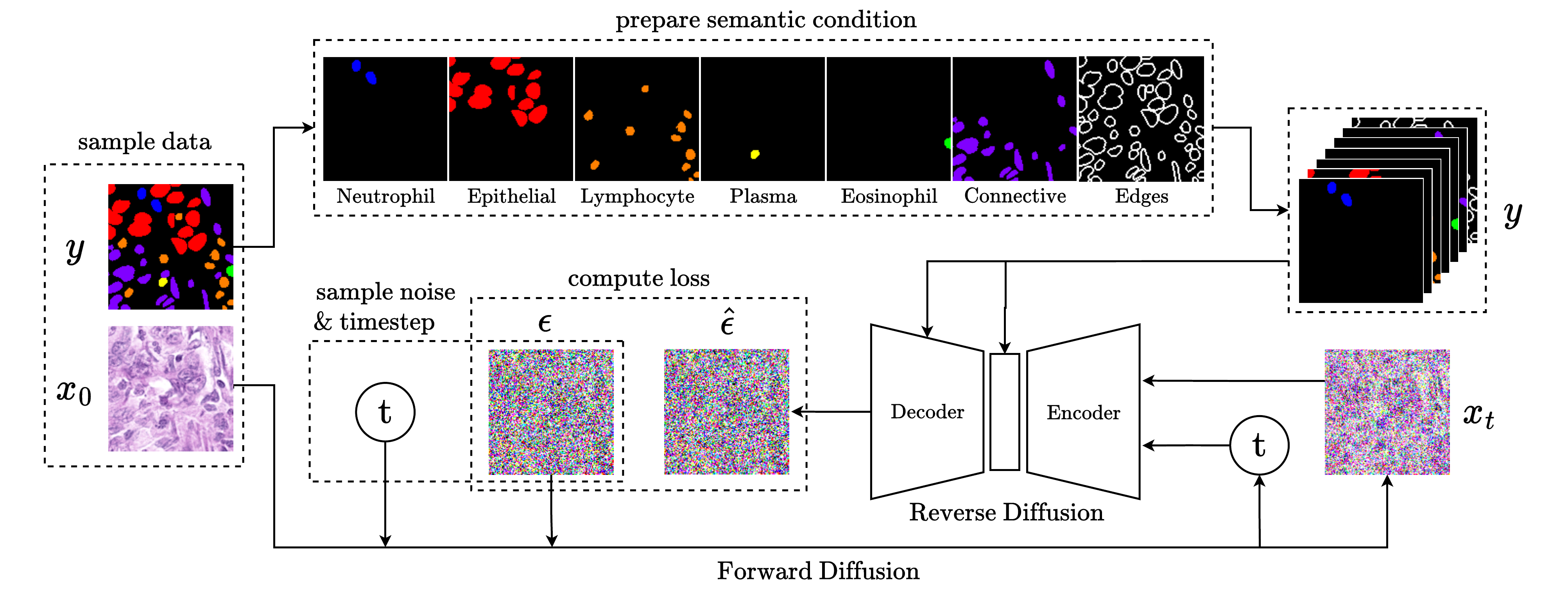}
\end{center}
\vspace{-0.2in}
   \caption{\textbf{NASDM training framework:} Given a real image $x_0$ and semantic mask $y$, we construct the conditioning signal by expanding the mask and adding an instance edge map. We sample timestep $t$ and noise $\epsilon$ to perform forward diffusion and generate the noised input $x_t$. The corrupted image $x_t$, timestep $t$, and semantic condition $y$ are then fed into the denoising model which predicts $\hat{\epsilon}$ as the amount of noise added to the model. Original noise $\epsilon$ and prediction $\hat{\epsilon}$ are used to compute the loss in~\eqref{eq:loss}.}
\vspace{-0.2in}
\label{fig:main}
\end{figure}

\subsection{Conditional Denoising Diffusion Probabilistic Model}
In this section, we describe the theory of conditional denoising diffusion probabilistic models, which serves as the backbone of our framework. A conditional diffusion model aims to maximize the likelihood $p_{\theta}(x_0 \mid y)$, where data $x_0$ is sampled from the conditional data distribution, $x_0 \sim q(x_0 \mid y)$, and $y$ represents the conditioning signal. A diffusion model consists of two intrinsic processes. The forward process is defined as a Markov chain, where Gaussian noise is gradually added to the data over $T$ timesteps as
\begin{equation}
    \begin{split}
        q(x_t \mid x_{t-1}) &= \mathcal{N}(x_{t}; \sqrt{1 - \beta_t} x_{t-1}, \beta_{t}\mathbf{I}),\\
        q(x_{1:T} \mid x_{0}) &= \prod^{T}_{t=1} q(x_t \mid x_{t-1}),
    \end{split}
\end{equation}
where $\{\beta\}_{t=1:T}$ are constants defined based on the noise schedule. An interesting property of the Gaussian forward process is that we can sample $x_{t}$ directly from $x_0$ in closed form. Now, the reverse process, $p_{\theta} (x_{0:T} \mid y)$, is defined as a Markov chain with learned Gaussian transitions starting from pure noise, $p(x_{T}) \sim \mathcal{N}(0, \mathbf{I})$, and is parameterized as a neural network with parameters $\theta$ as
\begin{equation}
    p_{\theta} (x_{0:T} \mid y) = p(y_T) \prod^{T}_{t=1} p_{\theta} (x_{t-1} \mid x_{t}, y).
\end{equation}
Hence, for each denoising step from $t$ to $t-1$,
\begin{equation}
    p_{\theta}(x_{t-1} \mid x_{t}, y) = \mathcal{N}(x_{t-1}; \mu_{\theta}(x_{t}, y, t), \Sigma_{\theta}(x_{t}, y, t)).
\end{equation}

It has been shown that the combination of $q$ and $p$ here is a form of a variational auto-encoder~\cite{kingma2013auto}, and hence the variational lower bound (VLB) can be described as a sum of independent terms, $L_{vlb} := L_{0} + ... + L_{T-1} + L_{T}$, where each term corresponds to a noising step. As described in Ho et al.~\cite{ho2020denoising}, we can randomly sample timestep $t$ during training and use the expectation $E_{t, x_0, y, \epsilon}$ to estimate $L_{vlb}$ and optimize parameters $\theta$. The denoising neural network can be parameterized in several ways, however, it has been observed that using a noise-prediction based formulation results in the best image quality~\cite{ho2020denoising}. Overall, our NASDM denoising model is trained to predicting the noise added to the input image given the semantic layout $y$ and the timestep $t$ using the loss described as follows:

\begin{equation} \label{eq:loss}
    L_{\text{simple}} = E_{t, x, \epsilon} \left[ \left\| \epsilon - \epsilon_{\theta}(x_t, y, t) \right\|_2 \right].
\end{equation}

Note that the above loss function provides no signal for training $\Sigma_{\theta} (x_t, y, t)$. Therefore, following the strategy in improved DDPMs~\cite{ho2020denoising}, we train a network to directly predict an interpolation coefficient $v$ per dimension, which is turned into variances and optimized directly using the KL divergence between the estimated distribution $p_{\theta}(x_{t-1} \mid x_t, y)$ and the diffusion posterior $q(x_{t-1} \mid x_t, x_0)$ as $L_{\text{vlb}} = D_{KL}(p_{\theta}(x_{t-1} \mid x_t, y) \parallel q(x_{t-1} \mid x_t, x_0))$. This optimization is done while applying a stop gradient to $\epsilon(x_t, y, t)$ such that $L_{\text{vlb}}$ can guide $\Sigma_{\theta}(x_t, y, t)$ and $L_{\text{simple}}$ is the main guidance for $\epsilon(x_t, y, t)$. Overall, the loss is a weighted summation of the two objectives described above as follows:
\begin{equation} \label{eq:objective}
    L_{\text{hybrid}} = L_{\text{simple}} + \lambda L_{\text{vlb}}.
\end{equation}

\subsection{Conditioning on Semantic Mask} \label{sec:cond_on_mask}
NASDM requires our neural network noise-predictor $\epsilon_{\theta}(x_t, y, t)$ to effectively process the information from the nuclei semantic map. For this purpose, we leverage a modified U-Net architecture described in Wang et al.~\cite{wang2022semantic}, where semantic information is injected into the decoder of the denoising network using multi-layer, spatially-adaptive normalization operators. As denoted in Fig.~\ref{fig:main}, we construct the semantic mask such that each channel of the mask corresponds to a unique nuclei type. In addition, we also concatenate a mask comprising of the edges of all nuclei to further demarcate nuclei instances.

\subsection{Classifier-free guidance}

To improve the sample quality and agreement with the conditioning signal, we employ classifier-free guidance~\cite{ho2022classifier}, which essentially amplifies the conditional distribution using unconditional outputs while sampling. During training, the conditioning signal, i.e., the semantic label map, is randomly replaced with a null mask for a certain percentage of samples. This leads to the diffusion model becoming stronger at generating samples both conditionally as well as unconditionally and can be used to implicitly infer the gradients of the log probability required for guidance as follows:
\begin{equation}
    \begin{split}
        \epsilon_{\theta} (x_t \mid y) - \epsilon_{\theta} (x_t \mid \emptyset) &\propto \nabla_{x_t} \log p(x_t \mid y) - \nabla_{x_t} \log p(x_t), \\
        &\propto \nabla_{x_t} \log p(y \mid x_t),
    \end{split}
\end{equation}
where $\emptyset$ denotes an empty semantic mask. During sampling, the conditional distribution is amplified using a guidance scale $s$ as follows:
\begin{equation}
    \hat{\epsilon}_{\theta}(x_t \mid y) = \epsilon_{\theta} (x_t \mid y) + s \cdot \left[ \epsilon_{\theta}(x_t \mid y) - \epsilon_{\theta} (x_t \mid \emptyset) \right].
\end{equation}





\section{Experiments}
\label{sec:experiments}
\begin{figure}[t]
\begin{center}
\includegraphics[width=\linewidth]{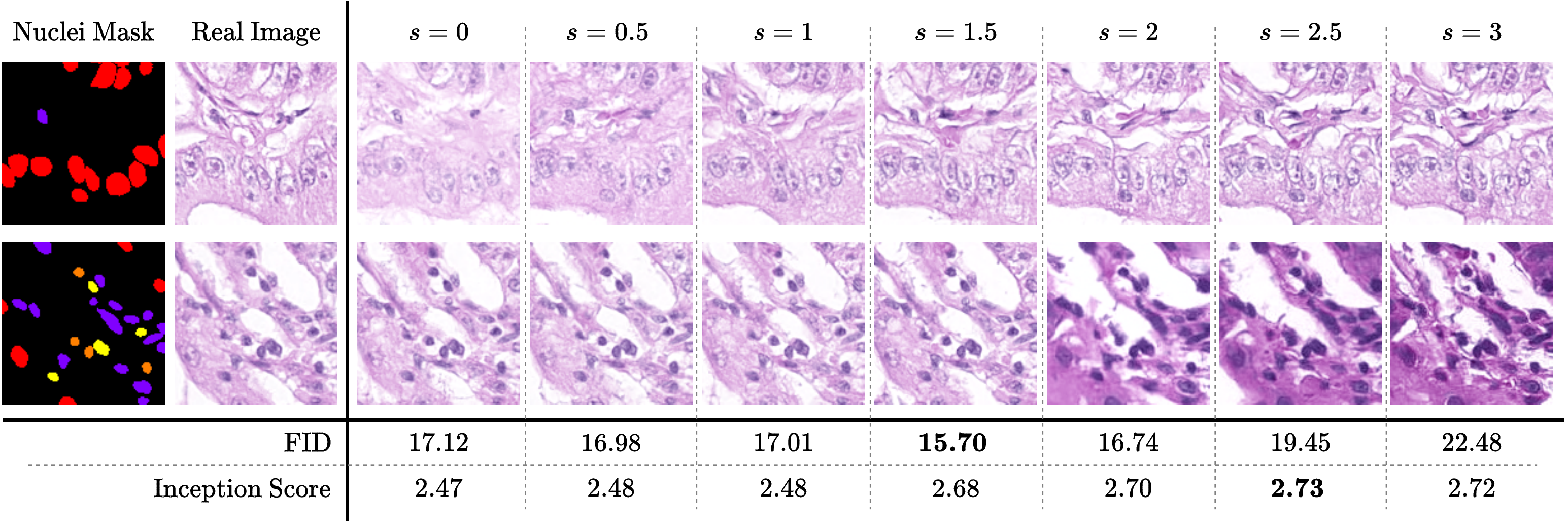}
\end{center}
\vspace{-0.2in}
   \caption{\textbf{Guidance Scale Ablation:} For a given mask, we generate images using different values of the guidance scale, $s$. The FID and IS metrics are computed by generating images for all masks in the test set at $20\times$ magnification.}
\vspace{-0.2in}
\label{fig:guidance}
\end{figure}

In this section, we first describe our implementation details and training procedure. Further, we establish the robustness of our model by performing an ablative study over objective magnification and classifier-guidance scale. We then perform quantitative and qualitative assessments to demonstrate the efficacy of our nuclei-aware semantic histopathology generation model. In all following experiments, we first synthesize images using the semantic masks of the held-out dataset at the concerned objective magnification. We then compute Fr\'echet  Inception Distance (FID) and Inception Score (IS) metrics between the synthetic and real images in the held-out set. 


\subsection{Implementation Details}
Our diffusion model is implemented using a semantic UNet architecture (Section~\ref{sec:cond_on_mask}), trained using the objective in~\eqref{eq:objective}. Following previous works~\cite{nichol2021improved}, we set the trade-off parameter $\lambda$ as $0.001$. We use the AdamW optimizer to train our model. Additionally, we adopt an exponential moving average (EMA) of the denoising network weights with $0.999$ decay. Following DDPM~\cite{ho2020denoising}, we set the total number of diffusion steps as $1000$ and use a linear noising schedule with respect to timestep $t$ for the forward process. After normal training with a learning rate of $1e-4$, we decay the learning rate to $2e-5$ to further finetune the model with a drop rate of $0.2$ to enhance the classifier-free guidance capability during sampling. The whole framework is implemented using Pytorch and trained on $4$ NVIDIA Tesla A100 GPUs with a batch-size of $40$ per GPU. Code will be made public on publication or request. 

\begin{table}[t]
\begin{center}
\setlength\tabcolsep{3pt}
\caption{\textbf{Quantitative Assesment:} We report the performance of our method using standard generative metrics Fr\'echet  Inception Distance (FID) metrics and Inception Score (IS) with the metrics reported in existing works. (-) denotes that the corresponding information was not reported in the original work.}
\label{tab:quant}
\begin{tabular}{lcccc}
    \toprule
    \textbf{Method} & \textbf{Tissue type} & \textbf{Conditioning} & \textbf{FID($\downarrow$)}  & \textbf{IS($\uparrow$)} \\ 
    \midrule
    BigGAN~\cite{brock2018large} & bladder & none & 158.4 & -           \\
    AttributeGAN~\cite{ye2021multi} & bladder & attributes & 53.6 & -           \\
    ProGAN~\cite{karras2017progressive} & glioma & morphology & 53.8 & 1.7           \\
    Morph-Diffusion~\cite{moghadam2023morphology} & glioma & morphology & 20.1 & 2.1 \\
    \midrule
    NASDM (Ours)  & colon & semantic mask & \textbf{15.7} & \textbf{2.7} \\
    \bottomrule
\vspace{-0.2in}
\end{tabular}
\end{center}
\end{table}

\subsection{Ablation over Guidance Scale ($s$)}
In this study, we test the effectiveness of the classifier-free guidance strategy. We consider the variant without guidance as our baseline. As seen in Figure~\ref{fig:guidance}, increase in guidance scale initially results in better image quality as more detail is added to visual structures of nuclei. However, with further increase, the image quality degrades as the model overemphasizes the nuclei and staining textures. 

\subsection{Ablation over Objective Magnification}

\begin{wraptable}{}{0.3\columnwidth}
\vspace{-0.3in}
\centering	
 \resizebox{0.3\columnwidth}{!}{%
    \begin{tabular}{ccc}
    \toprule
    \textbf{Obj. Mag.} &  \textbf{FID($\downarrow$)}  & \textbf{IS($\uparrow$)} \\ 
    \midrule
    10$\times$ & 38.1 & 2.3           \\
    20$\times$ & \textbf{15.7} & \textbf{2.7} \\
    \bottomrule
    \end{tabular}
}
\vspace{-0.3in}
\end{wraptable}

As described in Section~\ref{sec:data_process}, we generate patches at two different objective magnifications of $10\times$ and $20\times$. In this section, we contrast the generative performance of the models trained on these magnification levels respectively. From the table on right, we observe that the model trained at $20\times$ objective magnification produces better generative metrics. However, we hypothesize that the loss of expressiveness of the model at a lower magnification of $10\times$ could be because of the reduction in the training data at this magnification scale.

\begin{figure}[t] 
\begin{center}
\includegraphics[width=\linewidth]{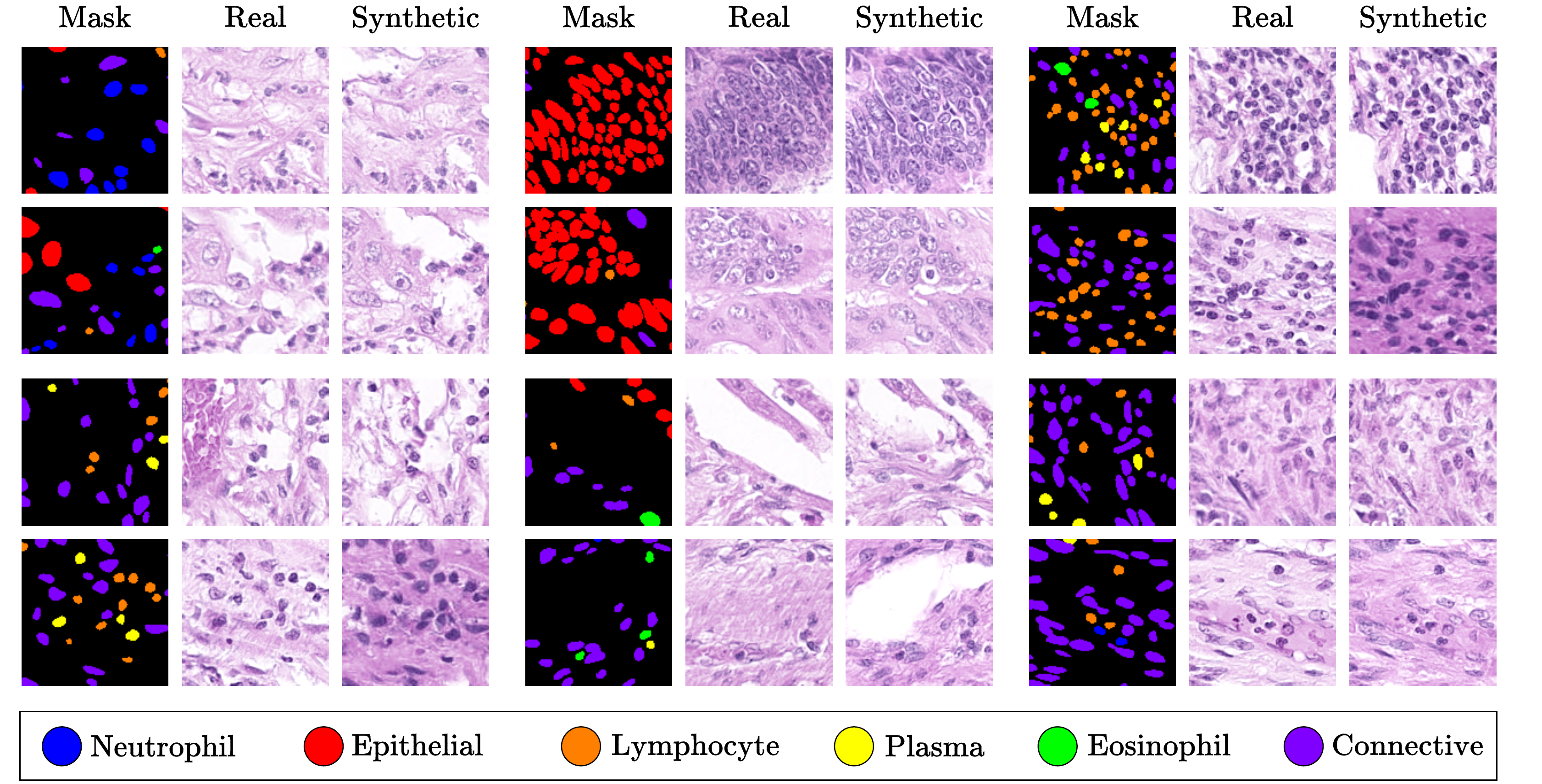}
\end{center}
\vspace{-0.2in}
   \caption{\textbf{Qualitative Analysis:} We generate synthetic images given masks with each type of nuclei in different environments to demonstrate the proficiency of the model to generate realistic nuclei arrangements. Legend at bottom denotes the mask color for each type of nuclei.}
\label{fig:qualitative}
\vspace{-0.2in}
\end{figure}

\subsection{Quantitative Analysis}
We compare the performance of our framework with existing histopathology generative models. To the best of our knowledge, ours is the only work that is able to synthesize histology images given a semantic mask, making a direct quantitative comparison tricky. However, the standard generative metric Fr\'echet Inception Distance (FID) measures the distance between distributions of generated and real images in the Inception-V3~\cite{kynkaanniemi2022role} latent space, where a lower FID indicates that the model is able to generate images that are very similar to real data. Therefore, we compare FID and IS metrics with the values reported in existing works~\cite{ye2021multi, moghadam2023morphology} (ref. Table~\ref{tab:quant}) in their own settings. We can observe that our method outperforms all existing methods including both GANs-based methods as well as the recently proposed morphology-focused generative diffusion model.

\subsection{Qualitative Analysis}
We now qualitatively discuss the proficiency of our model in generating realistic visual patterns in synthetic histopathology images (refer Fig.~\ref{fig:qualitative}). We demonstrate synthetic images with each type of nuclei in different environments. We can see that the model is able to capture convincing visual structure for each type of nuclei. In the synthetic images, we can see that the lymphocytes are accurately circular, while neutrophils and eosinophils have a more lobed structure. We also observe that the model is able to mimic correct nucleus-to-cytoplasm ratios for each type of nuclei. Epithelial cells are less dense, have a distinct chromatin structure, and are larger compared to other white blood cells. Due to their structure, epithelial cells are most difficult to generate in a convincing manner, however, we can see that model is able to capture the nuances well and generates accurate chromatin distributions. 




\vspace{-0.1in}
\section{Conclusion and Future Works}
\label{sec:discussion}
In this work, we present NASDM, a nuclei-aware semantic tissue generation framework. We demonstrate the model on a colon dataset and qualitatively and quantitatively establish the proficiency of the framework at this task. In future works, further conditioning on properties like stain-distribution, tissue-type, disease-type, etc. would enable patch generation in varied histopathological settings. Additionally, this framework can be extended to also generate semantic masks enabling an end-to-end tissue generation framework that first generates a mask and then synthesizes the corresponding patch. Further, future works can explore generation of patches conditioned on neighboring patches, as this enables generation of larger tissue areas by composing patches together. 


\bibliographystyle{splncs04}
\bibliography{egbib}
\end{document}